\documentclass[aps,prl,preprint,tightenlines,superscriptaddress,showpacs,byrevtex]{revtex4}
\usepackage{graphicx}
\usepackage{dcolumn}

\graphicspath{{ps}}

{\renewcommand{\thefootnote}{\fnsymbol{footnote}}

\begin{document}

\preprint{\vbox{ \hbox{   }
                 \hbox{BELLE-CONF-0628}
}}

\title{ \quad\\[0.5cm] Measurement of charmless $B$ Decays to $\eta K^*$ 
and $\eta \rho$ }

\affiliation{Budker Institute of Nuclear Physics, Novosibirsk}
\affiliation{Chiba University, Chiba}
\affiliation{Chonnam National University, Kwangju}
\affiliation{University of Cincinnati, Cincinnati, Ohio 45221}
\affiliation{University of Frankfurt, Frankfurt}
\affiliation{The Graduate University for Advanced Studies, Hayama} 
\affiliation{Gyeongsang National University, Chinju}
\affiliation{University of Hawaii, Honolulu, Hawaii 96822}
\affiliation{High Energy Accelerator Research Organization (KEK), Tsukuba}
\affiliation{Hiroshima Institute of Technology, Hiroshima}
\affiliation{University of Illinois at Urbana-Champaign, Urbana, Illinois 61801}
\affiliation{Institute of High Energy Physics, Chinese Academy of Sciences, Beijing}
\affiliation{Institute of High Energy Physics, Vienna}
\affiliation{Institute of High Energy Physics, Protvino}
\affiliation{Institute for Theoretical and Experimental Physics, Moscow}
\affiliation{J. Stefan Institute, Ljubljana}
\affiliation{Kanagawa University, Yokohama}
\affiliation{Korea University, Seoul}
\affiliation{Kyoto University, Kyoto}
\affiliation{Kyungpook National University, Taegu}
\affiliation{Swiss Federal Institute of Technology of Lausanne, EPFL, Lausanne}
\affiliation{University of Ljubljana, Ljubljana}
\affiliation{University of Maribor, Maribor}
\affiliation{University of Melbourne, Victoria}
\affiliation{Nagoya University, Nagoya}
\affiliation{Nara Women's University, Nara}
\affiliation{National Central University, Chung-li}
\affiliation{National United University, Miao Li}
\affiliation{Department of Physics, National Taiwan University, Taipei}
\affiliation{H. Niewodniczanski Institute of Nuclear Physics, Krakow}
\affiliation{Nippon Dental University, Niigata}
\affiliation{Niigata University, Niigata}
\affiliation{University of Nova Gorica, Nova Gorica}
\affiliation{Osaka City University, Osaka}
\affiliation{Osaka University, Osaka}
\affiliation{Panjab University, Chandigarh}
\affiliation{Peking University, Beijing}
\affiliation{University of Pittsburgh, Pittsburgh, Pennsylvania 15260}
\affiliation{Princeton University, Princeton, New Jersey 08544}
\affiliation{RIKEN BNL Research Center, Upton, New York 11973}
\affiliation{Saga University, Saga}
\affiliation{University of Science and Technology of China, Hefei}
\affiliation{Seoul National University, Seoul}
\affiliation{Shinshu University, Nagano}
\affiliation{Sungkyunkwan University, Suwon}
\affiliation{University of Sydney, Sydney NSW}
\affiliation{Tata Institute of Fundamental Research, Bombay}
\affiliation{Toho University, Funabashi}
\affiliation{Tohoku Gakuin University, Tagajo}
\affiliation{Tohoku University, Sendai}
\affiliation{Department of Physics, University of Tokyo, Tokyo}
\affiliation{Tokyo Institute of Technology, Tokyo}
\affiliation{Tokyo Metropolitan University, Tokyo}
\affiliation{Tokyo University of Agriculture and Technology, Tokyo}
\affiliation{Toyama National College of Maritime Technology, Toyama}
\affiliation{University of Tsukuba, Tsukuba}
\affiliation{Virginia Polytechnic Institute and State University, Blacksburg, Virginia 24061}
\affiliation{Yonsei University, Seoul}
  \author{K.~Abe}\affiliation{High Energy Accelerator Research Organization (KEK), Tsukuba} 
  \author{K.~Abe}\affiliation{Tohoku Gakuin University, Tagajo} 
  \author{I.~Adachi}\affiliation{High Energy Accelerator Research Organization (KEK), Tsukuba} 
  \author{H.~Aihara}\affiliation{Department of Physics, University of Tokyo, Tokyo} 
  \author{D.~Anipko}\affiliation{Budker Institute of Nuclear Physics, Novosibirsk} 
  \author{K.~Aoki}\affiliation{Nagoya University, Nagoya} 
  \author{T.~Arakawa}\affiliation{Niigata University, Niigata} 
  \author{K.~Arinstein}\affiliation{Budker Institute of Nuclear Physics, Novosibirsk} 
  \author{Y.~Asano}\affiliation{University of Tsukuba, Tsukuba} 
  \author{T.~Aso}\affiliation{Toyama National College of Maritime Technology, Toyama} 
% ...
\collaboration{The Belle Collaboration}

\noaffiliation

\begin{abstract}
We report branching fractions
and $CP$ asymmetries for 
$B \to \eta K^*$ and $B \to \eta \rho$ decays.
These results are obtained from a $414\,{\rm
fb}^{-1}$ data sample collected at   
the $\Upsilon(4S)$ resonance,
with the Belle detector at the KEKB asymmetric 
energy $e^+ e^-$ collider.
The branching fractions, in parts per million, of
$\eta K^{*0}$, $\eta K^{*+}$, $\eta \rho^0$,
and $\eta \rho^+$ are $15.9\pm 1.2 \pm 0.9$,
$19.7^{+2.0}_{-1.9} \pm 1.4$,
$0.84^{+0.56}_{-0.51} \pm 0.18$,
and $4.1^{+1.4}_{-1.3} \pm 0.34$, respectively.
We find no evidence for 
$CP$ asymmetries in these modes.
\end{abstract}
\maketitle

\tighten

{\renewcommand{\thefootnote}{\fnsymbol{footnote}}}
\setcounter{footnote}{0}
\section{Introduction} 
Charmless hadronic $B$ decays play an important role in 
understanding CP violation in the $B$ meson system.
Studies of $B \to \eta K^*$ and $B \to \eta \rho$~\cite{charge}
are important examples of such decays.
In the standard model (SM), Penguin (tree) diagrams are
expected to dominate in $B \to \eta K^*$ ($B \to \eta \rho$) decays.
The large branching fraction for $B \to \eta K^*$ compared to that for
for $B \to\eta K$ decay~\cite{CLEO,BELLEETA,BABARETA}
can be explained qualitatively in terms of the interference between 
non-strange and strange components of the $\eta$ meson, but are higher 
than recent theoretical predictions~\cite{ali,cheng,pqcd,rosner}.
In a similar vein, the larger measured branching fraction for 
charged ($B^+ \to \eta K^{*+}$) {\sl vs.} neutral ($B^0 \to \eta K^{*0}$) 
decays may suggest an additional SU(3)-singlet 
contribution~\cite{pqcd,rosner,hou} or constructive interference 
between SM penguin and tree amplitudes or 
between SM- and New Physics(NP)-penguin amplitudes.

In the standard model,
direct CP violation (DCPV) occurs in decays that involve 
two (or more) amplitudes that have different CP conserving and 
CP violating phases.  
The partial rate asymmetry can be written as
\begin{eqnarray}
{\mathcal A}_{CP}(B\to f) = {\Gamma({\bar B \to \bar f})
- \Gamma({B \to f}) \over \Gamma({\bar B \to \bar f})
+ \Gamma({B \to f})} 
  = { 2 |A_1| |A_2| \sin\Delta\delta \sin\Delta\phi \over
  	|A_1|^2 + |A_2|^2 + 2 |A_1| |A_2| \cos\Delta\delta \cos\Delta\phi} ,
\end{eqnarray}
where $f$ denotes a self-tagging final state, 
$B$ is either $B^+$ or $B^0$ meson, 
$\bar B$ and $\bar f$ are the conjugate states,
and $\Delta\delta$ ($\Delta\phi$) is the difference 
of the CP-conserving (CP-violating) phases between amplitudes $A_1$ and $A_2$.
The charge asymmetry will be sizeable when the two amplitudes are of
comparable strength with significant phase differences.
SM Penguin diagrams are expected to dominate in the
$B\to\eta K^{*}$ decays, while tree diagrams 
are expected to dominate in $B^+ \to \eta \rho^+$ decays. 
Therefore, DCPV in both cases should be 
small. However, SM- and NP-penguin diagrams may interfere to generate 
sizable direct CP violation.
\section{Data Set and Apparatus}  

This analysis is based on a data sample collected at the 
$\Upsilon$(4S) resonance with the Belle detector~\cite{BelleNIM} 
at the KEKB~\cite{kekb} accelerator.
The data sample corresponds to an integrated 
luminosity of $414\,{\rm fb}^{-1}$ and contains 
$449 \times 10^{6}$ $B \bar{B}$ pairs. 

The Belle detector is designed to measure charged particles and
photons with high efficiency and precision. 
Charged particle tracking is provided
by a silicon vertex detector (SVD) and a central drift chamber
(CDC) that surround the interaction region. 
The charged particle
acceptance covers the laboratory polar angle between $\theta=17^\circ$
and $150^\circ$, measured from the $z$ axis that is aligned 
anti-parallel to the positron beam. 
Charged hadrons are distinguished by combining the
responses from an array of silica aerogel Cherenkov counters
(ACC), a barrel-like array of 128 time-of-flight scintillation
counters (TOF), and $dE/dx$ measurements in the CDC. The combined
response provides $K/\pi$ separation of at least $2.5\sigma$ for
laboratory momentum up to 3.5~GeV/c. Electromagnetic showers are
detected in an array of 8736 CsI(Tl) crystals (ECL) located inside
the magnetic volume, which cover the same solid angle as the
charged particle tracking system. The 1.5-T magnetic field is
is contained via a flux return return that consists 
of 4.7 cm thick steel plates, interleaved with resistive plate 
counters used for tracking muons.
Two inner detector configurations are used. A 2.0 cm beampipe and 
a 3-layer silicon vertex detector are used for the first sample of
$152 \times 10^{6}$ $B \bar{B}$ pairs, while a 1.5 cm beampipe, 
a 4-layer silicon detector and a small-cell inner drift chamber
are used to record the remaining $297 \times 10^{6}$ $B \bar{B}$ 
pairs~\cite{SVD2}.

For Monte Carlo (MC) simulation study, 
the signal events, generic $b \to c$ decays and 
charmless rare $B$ decays
are generated with the EVTGEN~\cite{evtgen} event generator.
The continuum MC
events are generated with the $e^+e^- \to \gamma^* \to q \bar{q}$ 
process in the JETSET generator. 
The GEANT3~\cite{geant} package is used for detector simulation.
\section{Event Selection and Reconstruction}

Hadronic events are selected based on the charged track multiplicity
and total visible energy sum, which give an efficiency
greater than 99\% for $B\bar B$ events.
All primary charged tracks are required to satisfy track quality cuts  
based on their impact parameters relative to the run-dependent 
interaction point (IP): within $\pm 2\,{\rm cm}$ along the $z$ axis 
and within $\pm 1.5\, {\rm cm}$ in the transverse plane.
Particle identification (PID) is based on the likelihoods
${\mathcal L_{\rm K}}$/(${\mathcal L_\pi+L_{\rm K}}$)  
for charged kaons and pions, respectively.  
A higher value of ${\mathcal L_{\rm K}}$/(${\mathcal L_\pi+L_{\rm K}}$)  
indicates a more kaon-like particle. 
PID cuts are applied to all charged particles except pions 
from $K^0_S$ in this analysis. 
Unless explicitly specified, the PID cuts are 
${\mathcal L_{\rm K}}$/(${\mathcal L_\pi+L_{\rm K}}$) $> 0.6$ for kaons
and $< 0.4$ for pions. The PID efficiencies are $85$\% for
kaons and $89$\% for pions, while the 
fake rates are $8$\% for pions faking kaons and 
$11$\% for kaons faking pions. 
In forming $\pi^0$ candidates from photon pairs, the photon energies 
must exceed 50 MeV and the $\pi^0$ momentum in the center of mass (CM) frame 
must exceed 0.35 GeV/$c$. $K_S^0$ candidates are reconstructed from pairs 
of oppositely charged tracks whose invariant mass lies within $\pm 10\, 
{\rm MeV}/c^2$ of the $K_S^0$ meson mass.
We also require the vertex of the $K^0_S$  
to be well reconstructed and displaced from the interaction point, 
and the $K^0_S$ momentum direction be consistent 
with the $K^0_S$ flight direction.  
\subsection{$\eta$ Meson Reconstruction}
Candidate $\eta$ mesons are reconstructed through
$\eta \to \gamma \gamma$ and $\eta \to \pi^+\pi^- \pi^0$.
If one of the photons from the 
former $\eta$ decay mode could be paired with another photon with 
reconstructed $\gamma\gamma$ mass within $3\sigma$ of the $\pi^0$ meson 
mass, then the $\eta$ candidate is discarded.
We relax the PID requirement 
for charged pions from the latter $\eta$ decay mode to 
${\mathcal L_{\rm K}}$/(${\mathcal L_\pi+L_{\rm K}}$) $< 0.9$.
The momenta of the $\eta$ candidate is 
recalculated by applying the $\eta$ mass constraint 
for $B$ reconstruction. 
The $\eta \to \gamma\gamma$ candidates must satisfy $|\cos\theta^*| 
< 0.90$, where $\theta^*$ is the angle between the photon 
direction in the $\eta$ rest frame and the $\eta$ momentum in the CM frame, 
so as to suppress the soft photon combinatorial
background and $B \to K^* \gamma$ feed-across. 
The reconstructed mass resolutions are 12~MeV/$c^2$ for  
$\eta \to \gamma \gamma$ and 3.5~MeV/$c^2$ for $\eta \to \pi^+ \pi^- \pi^0$.  
\subsection{$K^*$, $\rho$ Meson Reconstruction}
$K^{*0}$ candidates are reconstructed from $K^-\pi^+$ and 
$K_S^0\pi^0$ pairs, while $K^{*+}$ mesons are reconstructed from $K^+\pi^0$ 
and $K_S^0\pi^+$ pairs.
Candidate $K^*$ mesons are required to have  
reconstructed masses within $\pm$75~MeV/$c^2$ of the nominal value.
Candidate $\rho^0$ ($\rho^+$) mesons are reconstructed from 
$\pi^-\pi^+$ ($\pi^0\pi^+$) pairs. Each combination is required to have 
a reconstructed mass within $\pm$150 MeV/c$^2$ of the nominal value.   
\subsection{$B$ Meson Reconstruction}
$B$ meson candidates are reconstructed from $\eta K^{*0}$, 
$\eta K^{*+}$, $\eta \rho^0$, and $\eta \rho^+$ combinations. They are 
characterized by the beam-constrained mass 
$M_{\rm bc}$ = $\sqrt{E^2_{\rm beam}-|P_B|^2}$ and the energy difference 
$\Delta E = E_B - E_{\rm beam}$, where $E_{\rm beam} = 5.29$~GeV, and 
$P_B$ and $E_B$ are the momentum and energy, respectively, of $B$ 
candidate in the $\Upsilon(4S)$ rest frame. 
We define the fit region in the $M_{\rm bc}$--$\Delta E$ plane as 
$M_{\rm bc} > 5.2\,{\rm GeV}/c^2$ and $|\Delta E| < 0.25\,{\rm GeV}$. We 
define the signal region as the overlap of the bands 
$M_{\rm bc} > 5.27\,{\rm GeV}/c^2$ and $|\Delta E| < 0.05\,{\rm GeV}$.

One candidate per event is required for all modes. The best candidate 
is chosen based on the sum of $\eta$ vertex mass constraint fit 
$\chi^2$ and $K^{*0}(\rho^0)$ vertex $\chi^2$ 
for $B\to \eta K^{*0}, K^{*0} \to K^+ \pi^-$ and $B \to \eta \rho^0$. 
For the $B^+ \to \eta K^{*+}$, $K^{*+} \to K^0 \pi^+$ mode, 
the reconstructed $K^{*+}$ mass $\chi^2$ and 
the $\eta$ vertex mass constraint fit 
$\chi^2$ are used.
For the remaining modes, the $\chi^2$ values of the $\eta$ and 
$\pi^0$ mass constraint fits are used.
\section{Background Suppression}

The dominant background for exclusive two-body
$B$ decays comes from the 
$e^+e^- \to \gamma^* \to q \bar{q}$ continuum ($q = u,\,d,\,s,\,c$),
which has a jet-like event topology compared to the spherical
$B\bar{B}$ events. Other major backgrounds involve feed-across
from other charmless $B$ decays. 
For decays involving the $\rho$ meson, an additional background 
arises from $b \to c$ decays. However, the impact of this background is 
small since the $M_{\rm bc}$ and $\Delta E$ distributions do not peak in 
the signal region.

\subsection{Continuum Background}

Signal and continuum events are distinguished by two means. First, 
we require $|\cos\theta_T| < 0.9$, where $\theta_T$ is defined as the 
angle between the $\eta$ daughter of a $B$ candidate and the thrust axis 
from all the particles in the event not associated with that $B$ 
candidate. This retains 90\% of signal and removes $\sim$56\% of 
continuum. Second, a probability density function ${\cal L}_s$ 
(${\cal L}_c$) for signal (continuum) is formed from two independent 
variables---$\cos\theta_B$, where $\theta_B$ is the polar angle of the 
$B$ candidate, and a Fisher discriminant~\cite{fisher} 
${\cal F} = \vec{\alpha}\cdot \vec{R}$ that combines seven 
event shape variables: $\cos\theta_T$, 
$S_{\perp}$ (the scalar sum of the transverse momenta of all particles 
outside a $45^{\circ}$ cone around the $B$ candidate direction divided 
by the scalar sum of the momenta), and the five modified Fox-Wolfram 
moments~\cite{fw} $R_2^{so}$, $R_4^{so}$, $R_2^{oo}$, $R_3^{oo}$, and 
$R_4^{oo}$. The Fisher discriminant's weight vector $\vec{\alpha}$ is 
determined by optimizing the separation between signal events and
continuum background using MC data; these Fox-Wolfram moments 
are used since they are not correlated with $M_{\rm bc}$. 
The likelihood ratio ${\cal LR} = {\cal L}_s / ({\cal L}_s + {\cal L}_c)$ 
is used to distinguish signal from continuum, since it peaks near 1 
for signal and near 0 for continuum.

The distribution of ${\cal LR}$ is found to depend somewhat on the 
event's $B$ flavor tagging quality parameter $r$~\cite{btag}, which ranges 
from zero for no flavor identification to unity for unambiguous flavor 
assignment. We partition the data into three $r$ regions, and determine 
the optimal cut on ${\cal LR}$ in each region by maximizing the significance 
$N_S/\sqrt{N_S+N_B}$ formed from the retained number of signal ($N_S$) and 
continuum background ($N_B$) events in MC samples.
A typical cut of ${\mathcal LR} > 0.5$
is $\sim 75$\% efficient for signal and removes $\sim 81$\% of the 
continuum background, while ${\mathcal LR} > 0.9$
is $\sim 38$\% efficient for signal and removes $\sim 97$\% 
of the continuum background.

\subsection{Feed-across and $b \to c$ Backgrounds}

A $B \to K^*(\rho) \gamma $ veto is applied to suppress 
the $K^*(\rho) \gamma $ feed-down for $B \to \eta K^*(\rho)$, 
$\eta \to \gamma \gamma$ decays.
The MC studies show that the background from $b \to c$ decays
and the feed-across from other charmless
$B$ decays are negligible for $B \to \eta K^*$ decays.
For $B \to \eta\rho$ decays, the background can be as large as 
4\% of the total yield (Table ~\ref{etarho_bg}). 
The contributions of these backgrounds are taken into account in the analysis.

\begin{table}[hb]
\begin{center}
\caption{Estimated yields from $b \to c$ ($N_{\rm bc}$), 
charmless $B$ decay ($N_{\rm r}$) backgrounds,
$\eta K^*$ feed-across ($N_{\rm feed}$) and 
actual yields from all sources ($N$) after the $\cos\theta_T$ 
and ${\cal LR}$ cuts.}
\begin{tabular}{lccc||c} \\ \hline \hline
Mode & $N_{\rm bc}$ & $N_{\rm r}$ & $N_{\rm feed}$ &$N$ \cr\hline
$\eta_{\gamma\gamma} \rho^0$ & $62$ & $81$ & $17$ & $2931$  \cr
$\eta_{\pi\pi\pi^0} \rho^0$ & $67$ & $27$ & $5$ & $1063$  \cr \hline
$\eta_{\gamma\gamma} \rho^+$ & $148$ & $74$ & $3$ & $4169$  \cr
$\eta_{\pi\pi\pi^0} \rho^+$ & $76$ & $22$ & $1$ & $1809$  \cr \hline
\hline
\end{tabular}
\label{etarho_bg}
\end{center}
\end{table}

\section{Analysis Procedure}

Signal yields are obtained using an extended unbinned
maximum likelihood (2-D ML) fit to the $M_{\rm bc}$ and $\Delta E$
distributions in the $M_{\rm bc}$--$\Delta E$ fit region that satisfy the 
$\cos\theta_T$ and ${\cal LR}$ requirements.

For $N$ input candidates, the likelihood is defined as
\begin{eqnarray}
L(N_S,N_B) = \frac{e^{-(N_S+N_B+N_{\rm bc}+N_{\rm r}+N_{\rm feed})}}{N!} 
\prod_{i=1}^{N}[N_{S} P_{S_i} + N_{B} P_{B_i} + N_{\rm bc} P_{{\rm bc}_i}
+N_{\rm r} P_{{\rm r}_i}+N_{\rm feed} P_{{\rm feed}_i}],
\end{eqnarray}
where $P_{S_i}$, $P_{B_i}$, $P_{{\rm bc}_i}$,
$P_{{\rm r}_i}$ and $P_{{\rm feed}_i}$ are the probability 
densities for event $i$ to be the signal, continuum, $b \to c$,
charmless $B$ decay and feed-across backgrounds 
for variables $M_{\rm bc}$ and $\Delta E$, respectively.  
Poisson statistics for $N_S$ and $N_B$, the extracted yields
for the signal and continuum background from the fit, are considered 
in this type of likelihood $L$. 
The yields $N_{\rm bc}$, $N_{\rm r}$, and $N_{\rm feed}$ are 
fixed from the MC analysis.

The continuum, $ b \to c$ and charmless B-decay
background $\Delta E$ PDF's are modeled by higher order
polynomial functions. The continuum and $b \to c$
background components in $M_{\rm bc}$ are
modeled by a smooth function~\cite{argus}. 
Due to the peaking behavior of $M_{\rm bc}$
in the signal region from charmless B decay
backgrounds, we use a sum of two bifurcated-Gaussian functions
to model the distributions. 
The bifurcated Gaussian combines the left half of a wide-resolution 
Gaussian with the right half of a narrow-resolution Gaussian, both having 
a common mean.
For $B \to \eta \rho$ decays, $M_{\rm bc}$ and $\Delta E$ distributions from 
$\eta K^*$ feed-across will behave like signal with a $\Delta E$ 
shift of $-50$ MeV. The PDF shape for each contributions
is determined by MC. 
The slope of the continuum-background $\Delta E$ polynomial and the 
parameters of the $M_{\rm bc}$ function are allowed to float in each fit.

For signal $\Delta E$, we used two bifurcated-Gaussian functions. 
The first accounts for $60$-$80$\% of the total area and the second, which 
has a larger width, is used to model the low-energy tail.
$M_{\rm bc}$ is weakly correlated with $\Delta E$, so we construct 
separate bifurcated Gaussians for $M_{\rm bc}$ in the three ranges 
$|\Delta E| < 0.05\,{\rm GeV}$, $0.05\,{\rm GeV} < |\Delta E| < 
0.1\,{\rm GeV}$, and $0.1\,{\rm GeV} < |\Delta E| < 0.25\,{\rm GeV}$.

For decays with more than one sub-decay process,
the final results are obtained
by fitting the sub-decay modes simultaneously  
with the expected 
efficiencies folded in and with the branching fraction
as the common output.
This is equivalent to summing
$\chi^2 = -2\ln(L)$
as a function of the branching fraction for
the sub-decay channels. 
The statistical significance ($\Sigma$) of the signal
is defined as $\sqrt{-2\ln(L_0/L_{\rm max})}$,
where $L_0$ and $L_{\rm max}$ denote
the likelihood values at zero signal events
and the best fit numbers, respectively.

The 90\% confidence level (C.L.) upper limit $x_{90}$ is 
calculated from the the equation:
\begin{eqnarray}
\frac{\int_0^{x_{90}} L(x) \,dx}{\int_0^\infty L(x) \,dx}
 = 90\% \,.
\end{eqnarray}
For this calculation of $x_{90}$, the likelihood function is 
modified to incorporate the systematic uncertainty.
\begin{figure}[htb]
\includegraphics[width=0.85\textwidth,height=0.47\textwidth]{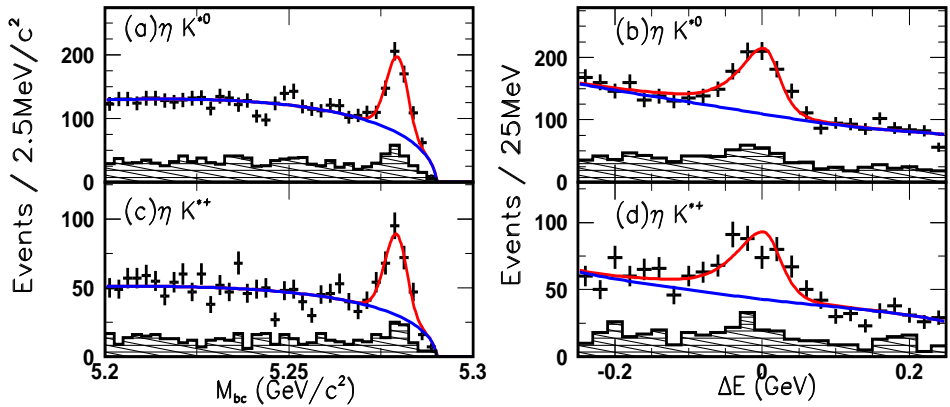}
\caption{Projections on $M_{\rm bc}$ (for the signal slice in $\Delta E$)
and $\Delta E$ (for the signal slice in $M_{\rm bc}$)
for $\eta K^{*0}$ (a,b) and $\eta K^{*+}$ (c,d) with the 
expected signal and background
function overlaid. The shaded area represents 
$\eta \to \pi^+ \pi^- \pi^0$ decays.}
\label{etafigmb}
\end{figure}
\section{Measurements of Branching Fractions}

The overall reconstruction efficiency $\epsilon$
is first obtained using MC samples
and then multiplied by PID efficiency corrections
obtained from data.
The PID efficiency correction is
determined by using $D^{*+} \to D^0 \pi^+$,
$D^0 \to K^- \pi^+$ data samples.
\begin{table}[htb]
\caption{Summary of results for each channel listed in the first column.
The measured signal yield ($N_S$), statistical significance ($\Sigma$),
reconstruction efficiency ($\epsilon$), 
total efficiency ($\epsilon_{tot}$) including the 
secondary branching fraction, 
and measured branching fractions are shown.
Uncertainties shown in second and sixth columns are statistical only.
The upper limit for $\eta \rho^0$ is at 90\% C.L.
\label{result}}
\vspace{0.1cm}
\begin{tabular}{lccccc}
\hline \hline
Mode & $N_S$ & $\epsilon$(\%) &
$\epsilon_{tot}$(\%) & $\Sigma$ & ${\mathcal B}(10^{-6})$\\
\hline\hline
$\eta_{\gamma\gamma} K^{*0}_{K^+\pi^-}$  &  $336.2^{+30.1}_{-29.2}$
              & $16.9$ & $4.4$ & 14.2 &
	      $16.9^{+1.5}_{-0.9}$ \\ 
$\eta_{\pi\pi\pi^0} K^{*0}_{K^+\pi^-}$  & $93.4^{+14.6}_{-13.8}$
              & $9.8$ & $1.5$  & 8.7 &
               $14.1^{+2.2}_{-2.1}$ \\
$\eta_{\gamma\gamma} K^{*0}_{K^0\pi^0}$  & $20.1^{+7.5}_{-6.7}$
              & $2.1$ & $0.27$ & 3.6 &
	      $16.7^{+6.3}_{-5.6}$ \\ 
$\eta_{\pi\pi\pi^0} K^{*0}_{K^0\pi^0}$  & $9.5^{+5.0}_{-4.2}$
              & $1.3$ & $0.098$  & 2.6 &
               $21.6^{+11.5}_{-9.7}$ \\\hline
$\eta K^{*0}$  & -
              & - & -  & $17.1$ &
               {$16.1\pm 1.2$} \\\hline\hline
$\eta_{\gamma\gamma} K^{*+}_{K^+ \pi^0}$ &  $79.8^{+16.1}_{-15.3}$
              & $6.7$ & $0.88$ & 6.1 &
	      $20.1^{+4.1}_{-3.9}$ \\ 
$\eta_{\pi\pi\pi^0} K^{*+}_{K^+ \pi^0}$ & $24.1^{+8.7}_{-7.9}$
              & $4.2$ & $0.32$ & $3.5$ &
	      $17.0^{+6.1}_{-5.6}$ \\ 
$\eta_{\gamma\gamma} K^{*+}_{K^0 \pi^+}$ & $120.3^{+16.2}_{-15.4}$
              & $4.5$ & $1.2$ & $10.1$ &
	      $22.6^{+3.1}_{-2.9}$ \\ 
$\eta_{\pi\pi\pi^0} K^{*+}_{K^0 \pi^+}$ & $29.2^{+7.3}_{-6.6}$
              & $2.6$ & $0.38$ & $6.2$ &
	      $17.0^{+4.8}_{-3.8}$ \\ \hline
$\eta K^{*+}$ & -
              & - & - & $13.8$ &
	      {$20.3^{+2.0}_{-1.9}$} \\ \hline\hline
$\eta_{\gamma\gamma} \rho^0$  & $19.5^{+11.3}_{-10.4}$
              & $8.9$ & $3.5$ & $2.1$ &
	      $1.25^{+0.73}_{-0.67}$ \\ 
$\eta_{\pi\pi\pi^0} \rho^0$  & $0.9^{+4.6}_{-3.9}$
              & $5.5$ & $1.2$  & $0.2$ &
               $0.17^{+0.84}_{-0.66}$ \\ \hline
$\eta \rho^0$  & -
              & - & -  & $1.6$ &
               $0.84^{+0.56}_{-0.51}$ ($< 1.5$) \\ \hline\hline
$\eta_{\gamma\gamma} \rho^{+}$ &$38.1^{+16.1}_{-15.2}$
              & $5.5$ & $2.2$ & $2.6$ &
	      $3.9^{+1.7}_{-1.6}$\\ 
$\eta_{\pi\pi\pi^0} \rho^+ $ & $15.8^{+8.9}_{-8.0}$
              & $3.50$ & $0.79$ & 2.1 &
	       $4.4^{+2.5}_{-2.2}$  \\ \hline
$\eta \rho^+$ & -
              & - & - & $3.4$ &
	      {$4.1^{+1.4}_{-1.3}$} \\ \hline\hline
\end{tabular}
\end{table}

Other MC efficiency corrections are determined
by comparing data and MC predictions
of other well-known processes.
The charged tracking efficiency correction is studied using
a high-momentum $\eta$ sample, comparing the ratio of
$\eta \to \pi^+ \pi^- \pi^0$ to $\eta \to \gamma\gamma$ between
data and MC. 
The same high-momentum $\eta$ sample is also used for $\pi^0$ 
reconstruction efficiency 
corrections by comparing the ratio of
$\eta \to \pi^0 \pi^0 \pi^0$ to $\eta \to \gamma\gamma$ between
the data and MC sample.
The $K^0_S$ reconstruction efficiency is verified by
comparing four $K^*(892)$ decay channels
($K^+\pi^-$, $K^+\pi^0$, $K^0_S\pi^+$, $K^0_S\pi^0$) in
inclusive $K^*$ and exclusive $B \to J/\psi K^*$ samples.
The ${\mathcal LR}$ cut efficiency correction is determined
using $B \to D \pi^+$ decays.
For $\eta$ and $K^*$ reconstruction and mass cuts,
we use the high-momentum $\eta$ and $K^*$ sample for
the efficiency correction studies.
\begin{figure}[htb]
\includegraphics[width=0.85\textwidth,height=0.47\textwidth]{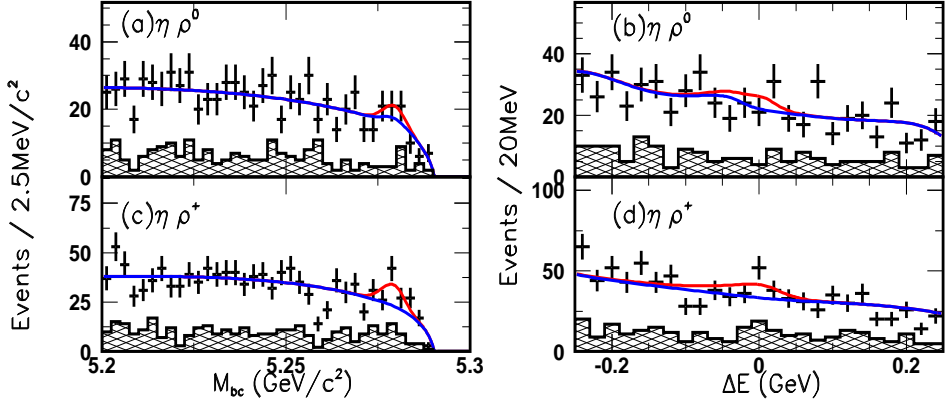}
\caption{Projections on $M_{bc}$ and $\Delta E$ from 2-D ML fitting results
for $\eta \rho^{0}$ (a,b) and $\eta \rho^{+}$ (c,d) with 
the expected signal and background
function overlaid. The shaded area represents 
$\eta \to \pi^+\pi^-\pi^0$ decays.}
\label{etarho}
\end{figure}
\begin{figure}[hb]
\includegraphics[width=0.85\textwidth]{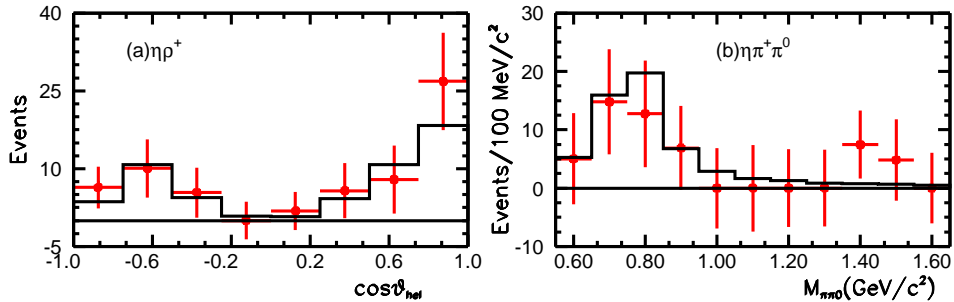}
\caption{Fitted yields vs (a) $\rho^+$ helicity,
(b)$\pi^+\pi^0$ invariance mass from $B\to\eta \rho^+$ decays. 
The overlaid histograms are expected distributions from MC and
normalized by the 2-D fit results.}
\label{etarhohelfig}
\end{figure}

The above studies show good agreement 
between the efficiencies in the data and MC 
sample at about the $2$\% level.
The PID, $\pi^0$, $\eta$ and $K^*$
reconstruction efficiency corrections are applied and
the systematic uncertainties are also obtained from the above studies.

The fitted signal yields and branching fractions are shown in 
Table~\ref{result}.
Several consistency checks are made, including
tighter ${\mathcal LR}$ cuts, 1-D ML $M_{bc}$, and $\Delta E$ fits,
and they are all shown to be consistent.
The total observed yields from the fits, with all sub-decay 
modes combined,
are $N_{\eta K^{*0}} = 459.2^{+34.6}_{-33.3}$ for
$B^0 \to \eta K^{*0}$,
$N_{\eta K^{*+}} = 253.4^{+25.5}_{-24.0}$
for $B^+ \to \eta K^{*+}$,
$N_{\eta \rho^0} = 20.4^{+12.2}_{-11.0}$ for
$B^0 \to \eta \rho^0$ and
$N_{\eta \rho^+} = 53.9^{+18.4}_{-17.1}$
for $B^+ \to \eta \rho^+$.
%
Figure \ref{etafigmb} shows the projections of 
of the data and the fits onto $M_{\rm bc}$ (for events in $\Delta E$ 
signal slice) and $\Delta E$ (for events in the $M_{\rm bc}$ signal slice) 
for the $B \to \eta K^*$ decays, while Fig.~\ref{etarho}
shows the corresponding projections for the $B \to \eta \rho$ decays. 
For $B^+ \to \eta \rho^+$ decays, where a clear excess is seen, 
we examine the properties of the $\rho^+$ candidates.
Although statistically limited,
clear $\rho^+$ mass peaks and a
polarized $\cos\theta_{\rm hel}$ distribution are 
observed (Fig.~\ref{etarhohelfig}) 
and are consistent with the expectation from $\eta \rho^{+}$
with no significant non-resonant $B^+ \to \eta \pi^+\pi^0$.
\begin{figure}[htb]
\includegraphics[width=0.85\textwidth]{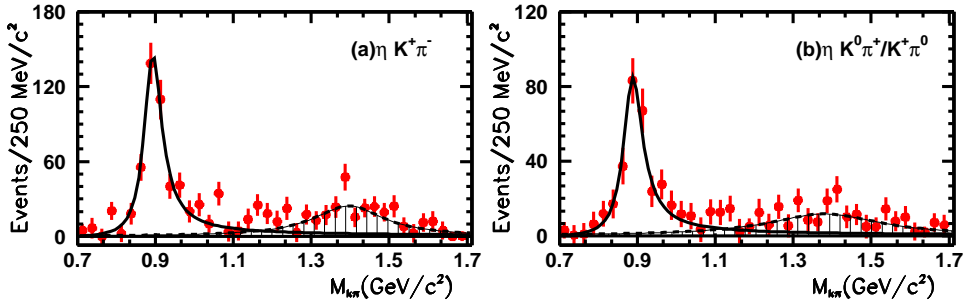}
\caption{Fitted yields vs the $K\pi$ invariant mass for the (a) $K^{*0}$
and (b) $K^{*+}$ modes. Overlaid functions are results from fitting
with a $K^*$ line shape and a higher resonance. Only the parameters
of the higher resonance are free during the fit.}
\label{kpifit}
\end{figure}

\section{Systematic Error}

Systematic errors arise from
efficiency corrections and fitting.
The main sources of uncertainties in the efficiency 
corrections are from
the reconstruction of low-momentum charged tracks,
low-energy photon finding, and ${\mathcal LR}$ cut efficiency,
each at a level of a few per cent.
The systematic errors include contributions of
$1$\% for ${\mathcal LR}$ cuts, $1$\% per reconstructed charged 
particle, $0.5$\%
for each charged particle identification, 
$4$\% for $\pi^0$ reconstruction, $4.5$\% for $K^0_S$ reconstruction,
and $2$\% for $\eta$ reconstruction with $\eta \to \gamma \gamma$.
To evaluate a possible non-resonant or higher resonance 
contribution in the
$K^*$ region, we perform a 2D ML fit within different $K\pi$ invariant
mass region from $0.7$~GeV/$c^2$ to $1.72$~GeV/$c^2$. 
Figure \ref{kpifit} shows our results with a $K^*$ line shape and 
a higher resonance overlaid with the $K^*$ area fixed.
Based on this study, the non-resonant $K\pi$ contributions
are $1.7 \pm 0.5$\% for $\eta K^{*0}$ and $3.7 \pm 2.0$\%
for $\eta K^{*+}$ decays. These corrections will be applied to
the final branching fraction measurements of $B \to \eta K^*$.
\begin{figure}[htb]
\includegraphics[width=0.85\textwidth]{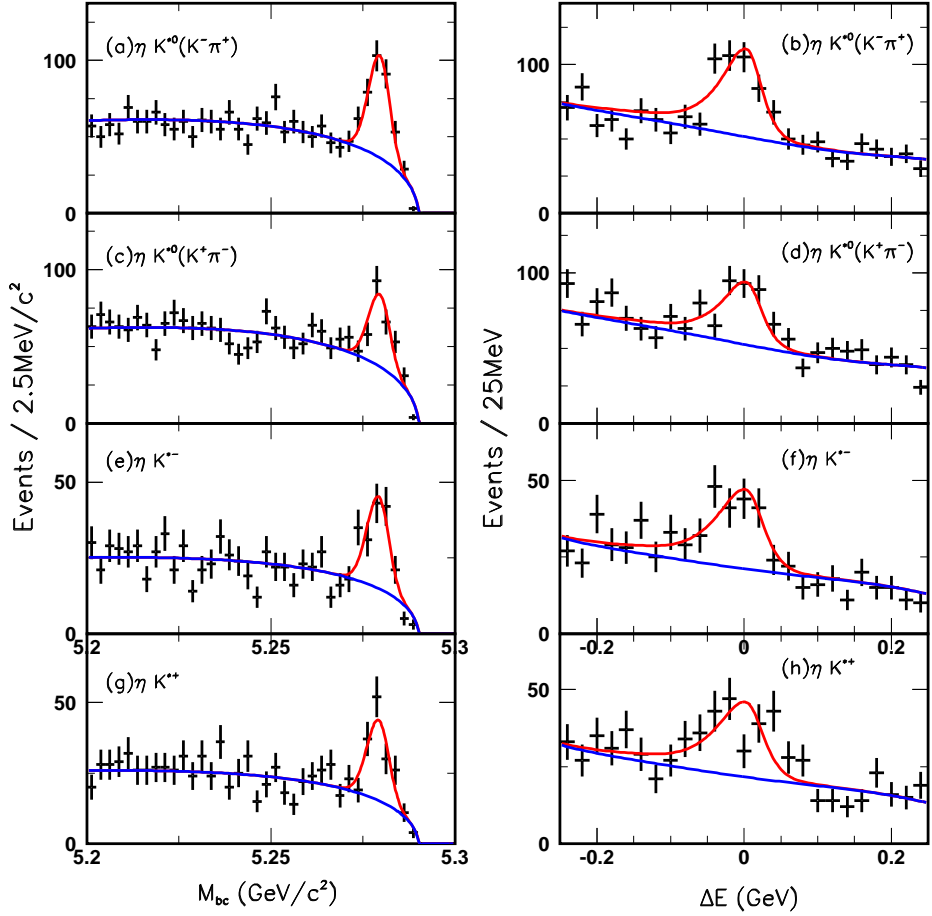}
\caption{Projection on $M_{\rm bc}$(left) and $\Delta E$(right) 
from 2-D ML fitting results for $\eta K^*$ with the expected 
signal and background function overlaid. 
Figures (a,b,e,f) are for $\bar b$
decays and (c,d,g,h) are for $b$ decays.}
\label{acpfig}
\end{figure}
Due to the limited statistics for 
$B^+ \to \eta \rho^+$ decays, 
a larger systematic error for the non-resonant or higher resonance
contributions is assigned with no corrections applied.

We use $B^0 \to {\bar D^0} \pi^+$ decays
to estimate the uncertainties in the signal
function PDF's used for fitting $M_{\rm bc}$ and $\Delta E$ 
by comparing the mean and the width of the $M_{\rm bc}$ and
$\Delta E$ distributions between the $B^0 \to {\bar D^0} \pi^+$ 
data and the MC sample.
For MC estimated $b \to c$ and charmless B decay backgrounds,
we vary the estimated yields by $50$\% and  
refit the data. The difference in the fit yields from the nominal values 
contributes to the systematic error. 
The overall relative systematic errors are
$5.8$\% for $\eta K^{*0}$,$7.2$\% for $\eta K^{*+}$,
$22$\% for $\eta \rho^0$ and $8.4$\% for $\eta \rho^+$. 
%

\section{$\mathcal A_{CP}$ Measurements}
We measure $\mathcal A_{CP}$ in the
decays of $B \to \eta K^*$ and $B^+ \to \eta \rho^+$.
Due to the wrong-tag fraction $w$, the value of $\mathcal A_{CP}$
is not the same as the measured values $\mathcal A^{obs}_{CP}$.
Their correlation is ${\mathcal A^{obs}_{CP}} = (1-2w){\mathcal A_{CP}}$.
In the decay modes we study, only those in which $\mathcal A_{CP}$ values
are determined by low momentum charged pions will
have a non-negligible $w$.
The wrong-tag fractions for $K^{*+} \to K^0\pi^+$ is $\sim 1.5$\%
for $\eta K^{*+}$ decays and $\sim 2.0$\% for $\eta \rho^+$ decays.
Other decays have $w < 0.1$\%. Since the result for
$\eta K^{*+}$ is obtained from a simultaneous fit to all four sub-decay modes
with roughly equal statistics for $K^{*+} \to K^+ \pi^0$
and $K^{*+} \to K^0 \pi^+$, 
the wrong-tag effect for $\mathcal A_{CP}$
should be $< 0.7$\%. Thus, there are no wrong-tag fraction corrections for
$\eta K^*$ decays. For $\eta \rho^+$, 2\% is assigned for $w$.

To incorporate the CP asymmetry in the fit, the coefficients of the 
signal and continuum background PDFs in the likelihood are modified: 
$N_S \to {1\over 2}N_S(1 - q{\cal A}_{CP}^{\rm obs})$ and 
$N_B \to {1\over 2}N_B(1 - q{\cal A}_{CP,qq})$,
where $q = +1 (-1)$ for a $B(\bar B)$ meson tag and 
$\mathcal A^{obs}_{cp}$,$\mathcal A_{cp,qq}$ are the
$\mathcal A_{cp}$ outputs for signal and 
continuum respectively.
The results are ${\mathcal A^{obs}_{cp}}(\eta K^{*0})=0.17 \pm 0.08$,
${\mathcal A^{obs}_{cp}}(\eta K^{*+})=0.03 \pm 0.10$ and
${\mathcal A^{obs}_{cp}}(\eta \rho^+)=-0.04^{+0.34}_{-0.32}$.
Figure \ref{acpfig} shows the projections of 
of the data and the fits onto $M_{\rm bc}$ (for events in $\Delta E$ 
signal slice) and $\Delta E$ (for events in the $M_{\rm bc}$ signal slice) 
for the $B \to \eta K^*$ decays.
%
%
%

Since the systematic errors in $\eta$ reconstruction and
the number of $B\bar{B}$ events cancel in the ratio,
the systematic uncertainty of $\mathcal {A_{CP}}$
is mainly from the asymmetry for PID of charged kaons 
and the fitting PDF's.
The efficiency asymmetry for the PID of charged kaons is $0.01$
in absolute value.

\section{Discussion and Conclusion}

In summary, we report measurements of the exclusive two-body 
charmless hadronic $B \to \eta K^*$ and $B \to \eta \rho$ decays 
with high statistics.
Our results confirm that the branching fractions for
$B^0\to \eta K^{*0}$ and $B^+\to \eta K^{*+}$ are large
and are consistent with previous measurements~\cite{BABARETA,babar1}.
The branching fractions obtained are
%
${\mathcal B}(B\to\eta K^{*0}) =
(15.9 \pm 1.2 \pm 0.9)\times 10^{-6}$, and
${\mathcal B}(B\to\eta K^{*+}) =
(19.7^{+2.0}_{-1.9}\pm 1.4)\times 10^{-6}$,
where the first error is statistical and the second systematic.
Our measurements imply that the branching fraction of
$B^+ \to \eta K^{*+}$ is 1$\sigma$ higher than 
$B^0 \to \eta K^{*0}$, which may constrain the contribution of the
flavor singlet penguin amplitude as suggested by various
theorists~\cite{hou,rosner}.
Large signals of $B^+ \to \eta \rho^+$ have been
observed  with the branching fraction
$\mathcal B$($B\to\eta \rho^+$)=
$(4.1^{+1.4}_{-1.3}\pm 0.34)\times 10^{-6}$.
The branching fractions and 90\% C.L. for $B^0 \to \eta \rho^0$ decays are
$\mathcal B$($B^0\to\eta \rho^0$)=
$(0.84^{+0.56}_{-0.51}\pm 0.18)\times 10^{-6}$
($< 1.9 \times 10^{-6}$).

We also search for direct CP asymmetry in $B \to \eta K^*$ and 
$B^+ \to \eta \rho^+$,
and final results are consistent with no asymmetry.
The results are: 
${\mathcal A_{cp}}(\eta K^{*0}) = 0.17 \pm 0.08 \pm 0.01,
{\mathcal A_{cp}}(\eta K^{*+}) =  0.03 \pm 0.10 \pm 0.01$, and
${\mathcal A_{cp}}(\eta \rho^+) = -0.04^{+0.34}_{-0.32}\pm 0.01$. 
\section{Acknowledgments} 
We thank the KEKB group for the excellent operation of the
accelerator, the KEK cryogenics group for the efficient
operation of the solenoid, and the KEK computer group and
the National Institute of Informatics for valuable computing
and Super-SINET network support. We acknowledge support from
the Ministry of Education, Culture, Sports, Science, and
Technology of Japan and the Japan Society for the Promotion
of Science; the Australian Research Council and the
Australian Department of Education, Science and Training;
the National Science Foundation of China and the Knowledge
Innovation Program of the Chinese Academy of Sciencies under
contract No.~10575109 and IHEP-U-503; the Department of
Science and Technology of India;
the BK21 program of the Ministry of Education of Korea,
the CHEP SRC program and Basic Research program
(grant No.~R01-2005-000-10089-0) of the Korea Science and
Engineering Foundation, and the Pure Basic Research Group
program of the Korea Research Foundation;
the Polish State Committee for Scientific Research;
the Ministry of Science and Technology of the Russian
Federation; the Slovenian Research Agency;  the Swiss
National Science Foundation; the National Science Council
and the Ministry of Education of Taiwan; and the U.S.\
Department of Energy.


\begin{thebibliography}{99}
\bibitem{charge}{Throughout this paper, the inclusion of charge conjugate 
        mode decays is implied unless otherwise stated. 
}
\end{thebibliography}
\end{document}